\documentclass[prl,aps,twocolumn,showpacs,superscriptaddress]{revtex4-1}
\usepackage{graphicx}
\usepackage{dcolumn}
\usepackage{bm}
\usepackage{float}
\begin{document}


\title{First-Principles Determination of the Structure of Magnesium Borohydride}

\author{Xiang-Feng Zhou}
\email{xfzhou@nankai.edu.cn/zxf888@163.com}
\affiliation{School of Physics and Key Laboratory of Weak-Light Nonlinear Photonics, Nankai University, Tianjin 300071, China}
\affiliation{Department of Geosciences, Department of Physics and Astronomy, Stony Brook University, Stony Brook, New York 11794, USA}

\author{Artem R. Oganov}
\affiliation{Department of Geosciences, Department of Physics and Astronomy, Stony Brook University, Stony Brook, New York 11794, USA}
\affiliation{Geology Department, Moscow State University, Moscow 119992, Russia}

\author{Guang-Rui Qian}
\affiliation{Department of Geosciences, Department of Physics and Astronomy, Stony Brook University, Stony Brook, New York 11794, USA}

\author{Qiang Zhu}
\affiliation{Department of Geosciences, Department of Physics and Astronomy, Stony Brook University, Stony Brook, New York 11794, USA}


\begin{abstract}
\noindent The energy landscape of Mg(BH$_4$)$_2$ under pressure is explored by \textit{ab initio} evolutionary calculations. Two new tetragonal structures, with space groups $P\bar{4}$ and $I4_1/acd$ are predicted to be lower in enthalpy, by 15.4 kJ/mol and 21.2 kJ/mol, respectively, than the earlier proposed $P4_2nm$ phase. We have simulated X-ray diffraction (XRD) spectra, lattice dynamics, and equations of state (EOS) of these phases. The density, volume contraction, bulk modulus, and the simulated XRD patterns of $I4_1/acd$ and $P\bar{4}$ structures are in excellent agreement with the experimental results.
\end{abstract}



\maketitle

Light weight metal borohydrides have recently received much attention owing to their high gravimetric and volumetric hydrogen densities compared to other complex hydrides \cite{R01,R02,R03,R04}. Of these, magnesium borohydride, Mg(BH$_4$)$_2$, is a prominent lightweight solid-state hydrogen storage material with a theoretical hydrogen capacity of 14.8 wt \%. Hence it has been extensively studied as a template for developing novel hydrogen-storage solutions. Based on the experimental data, the ground state $\alpha$ and $\beta$ phases, have been assigned space groups $P6_122$ (330 atoms per unit cell) and $Fddd$ (704 atoms/cell), which have unexpectedly complex crystal structures \cite{R05,R06,R07,R08,R09,R10}. There had been disputes between experimentalists and theoreticians regarding  the nature of the ground-state structure of Mg(BH$_4$)$_2$ \cite{R11,R12,R13,R14}. To improve the reversible hydrogen absorption or desorption kinetics or get new metastable polymorphs, a practical method is to search for possibility to stabilize the high-pressure phase of Mg(BH$_4$)$_2$ at ambient pressure by advanced theoretical or experimental methods. For example, $\gamma$-Mg(BH$_4$)$_2$ is one of the hydrogen-richest solids, reported to be capable of storing guest species, such as hydrogen \cite{R14}. Most recently, new $\delta$, $\delta'$, and $\epsilon$ phases of Mg(BH$_4$)$_2$ were successfully synthesized under pressure \cite{R14}. Many of them turned out to retain their structure upon decompression to ambient conditions. Crystal structures of $\gamma$ and $\delta$ phases were, apparently convincingly, resolved using powder x-ray diffraction data obtained with synchrotron radiation \cite{R14}. Unexpectedly, our phonon calculations showed the $P4_2nm$ structure ($\delta$ phase) to be dynamically unstable at ambient pressure, which means that the exact crystal structure of $\delta$ phase is still unresolved, even for such a simple structure only with 22 atoms per cell, and still less for the unknown $\delta'$ and $\epsilon$ phases \cite{R14}. Therefore, the polymorphism and phase diagram of this important compound require further investigation.

Structure searching for ground-state phases of Mg(BH$_4$)$_2$ were performed with the \textit{ab initio} evolutionary algorithm \textsc{uspex} \cite{R15,R16,R17}. This method has been successfully applied to a wide range of problems \cite{R18,R19}. When nothing is known about the crystal structure, separate searches are usually done with increasing number of formula units in the unit cell -- as long as computational resources allow and until the solution is found. We found structure solutions already in searches with 2 formula units (22 atoms/cell) and 4 formula units (44 atoms/cell), which yielded structures fully explaining experimental results. We did structure prediction at 0, 2, 5, 10, 15, and 20 GPa. All structures were relaxed by the \textsc{vasp} code \cite{R20}. During structure searches, we used BH$_4$ groups as whole units with fixed bond connectivity. All experimentally known structures contain these nearly rigid units at pressures below 22 GPa \cite{R06}. This was done with the \textit{Z-matrix} representation for the BH$_4$ complex anions, keeping the bond connectivity within the BH$_4$ tetrahedron, and fully relaxing the bond lengths and angles. This approach allowed us to focus searches on chemically interesting structures and reduced the search space dramatically \cite{R17}. By the way, the presence of the same nearly rigid BH$_4$ groups, connected to Mg atoms by weak bonds implies that zero-point energies of different structures will be nearly identical and thus can be safely neglected when comparing energetics of different structures; our calculations confirm this. We employed projector augmented wave method to describe the core electrons and the generalized gradient approximation of Perdew, Burke, and Ernzernhof (PBE) for exchange and correlation. A cutoff energy of 600 eV, and a Monkhorst-Pack Brillouin zone sampling grid with the resolution of $2 \pi \times 0.04$ \AA$^{-1}$ was used. To examine the importance of long-range dispersion interactions on this compound's thermodynamic stability, the semiempirical dispersion-correction method was applied to the ground state structures \cite{R21}. Phonon dispersion curves were calculated by using the supercell method as implemented in the \textsc{phonopy} code \cite{R22,R23}. The powder XRD patterns were simulated using the \textsc{reflex} software.

\begin{figure}[h]
\begin{center}
\includegraphics[width=8cm]{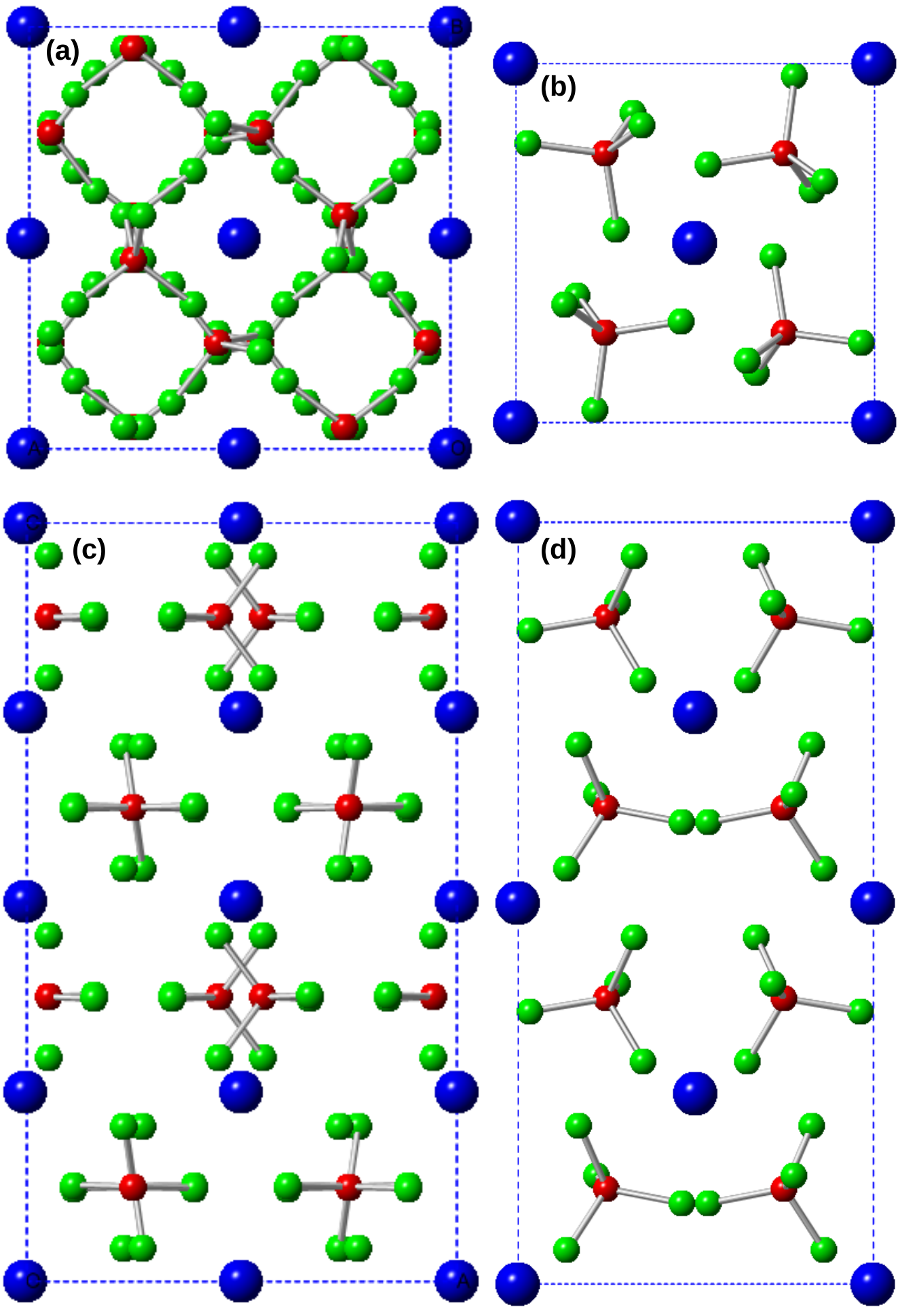}
\caption{%
(Color online) (a), (c) Projections of the $I4_1/acd$ structure along [001] and [010] directions, (b), (d) projections of the $P\bar{4}$ structure along [001] and [010] directions. Large spheres are Mg atoms, small spheres -- B (dark or red) and H (light or green) atoms.}
\end{center}
\end{figure}

In our calculations, we found $F222$ Mg(BH$_4$)$_2$ as the most stable phase at ambient pressure, in good agreement with previous theoretical results \cite{R12,R13}. The tetragonal $I4_1/acd$ and trigonal $P\bar{3}m1$ phases are found to be the most stable ones in structure searches at 2-5 GPa and 10-20 GPa, respectively. Interestingly, within the whole pressure range (up to 20 GPa), we did not find the $P4_2nm$ structure proposed by Filinchuk \textit{et al}. for the $\delta$ phase \cite{R14}, but instead found the $I4_1/acd$ phase with four formula units (44 atoms) per conventional cell and $P\bar{4}$ phase with 2 formula units per cell at pressures below 5 GPa (see Fig. 1). Given that the $P4_2nm$ structure is dynamically unstable at ambient pressure, and based on our enthalpy calculations, we hypothesized that the $I4_1/acd$ and $P\bar{4}$ structures might correspond to the experimentally observed $\delta$ and $\delta'$ phases. Further investigation confirmed this hypothesis, as we will show below.

\begin{table}
\caption{Lattice constants, density ($\rho$), bulk modulus (B$_0$) and its pressure derivative ($B'$), and the total energy of the polymorphs of Mg(BH$_4$)$_2$. The energy difference ($\Delta E$) including van der Waals interactions ($\Delta E'$) is relative to the $P4_2nm$ structure. Some experimental values (from Ref. 14) are also listed for comparison.}
\begin{tabular}{lccccc}
\hline\hline
Symmetry & $I4_1/acd$ & $I4_1/amd$ & $P\bar{4}$ & $P4_2nm$ & $P4_2nm$ \\
         &            &            &            & (Theory) & (Expt) \\
\hline
$a$ ({\AA}) & 7.69 &  8.32 & 5.58 & 5.79 & 5.44 \\
$b$ ({\AA}) & 7.69 &  8.32 & 5.58 & 5.79 & 5.44 \\
$c$ ({\AA}) & 12.30 & 10.52 & 5.99 & 5.73 & 6.15 \\
$\rho$ ($gcm^{-3}$) & 0.986 & 0.984 & 0.963 & 0.933 & 0.987\\
B$_0$ (GPa) & 24.0 & 31.0 & 31.7 & 28.5 & 28.5 \\
$B'$        & 4.3 & 3.6 & 3.6 & 3.6 & 5.8  \\
$\Delta E$  (kJ/mol) & -17.0 & -15.5 & -13.6 & 0  &\\
$\Delta E'$  (kJ/mol) & -21.2 & -19.3 & -15.4 & 0  &\\
\hline\hline
\end{tabular}
\end{table}

Lattice constants of the most interesting candidate structures are listed in Table I. The calculated zero-pressure lattice constant ($a$ = $b$ = $c$ = 15.79 {\AA}) of the $\gamma$ phase is in excellent agreement with the experimental value (15.76 {\AA}) \cite{R14}, which gives a benchmark of the typical accuracy to expect of DFT simulations for this system. For the $I4_1/acd$ structure, the Mg atom occupies the crystallographic $8b$ site at (0.5, 0, 0), the B atom occupies the $16e$ site at (0.533, 0.25, 0.375), and the H atoms are at the $32g$ sites with coordinates (0.439, 0.247, 0.295) and (0.627, 0.123, 0.369). For the $P\bar{4}$ structure, the Mg atoms occupies the $1a$ site at (0, 0, 0) and $1d$ site at (0.5, 0.5, 0.5), the B atom occupies at the $4h$ site at (0.25, 0.25, 0.25), and the H atoms are at the $4h$ sites with coordinates (0.459, 0.282, 0.210), (0.139, 0.327, 0.089), (0.223, 0.038, 0.286), and (0.174, 0.354, 0.416). From Table I, one can see that relaxing the experimental $P4_2nm$ structure, one gets unexpectedly large changes in the lattice constants -- so large that, in fact, the relaxed lattice constants of the $P\bar{4}$ structure are the closest match to the experimental ones \cite{R14}. One has to keep in mind that what is called the ``experimental" cell parameters in many cases is a non-unique result of indexing powder XRD spectra, and this is the case here.

\begin{figure}[h]
\begin{center}
\includegraphics[width=8cm]{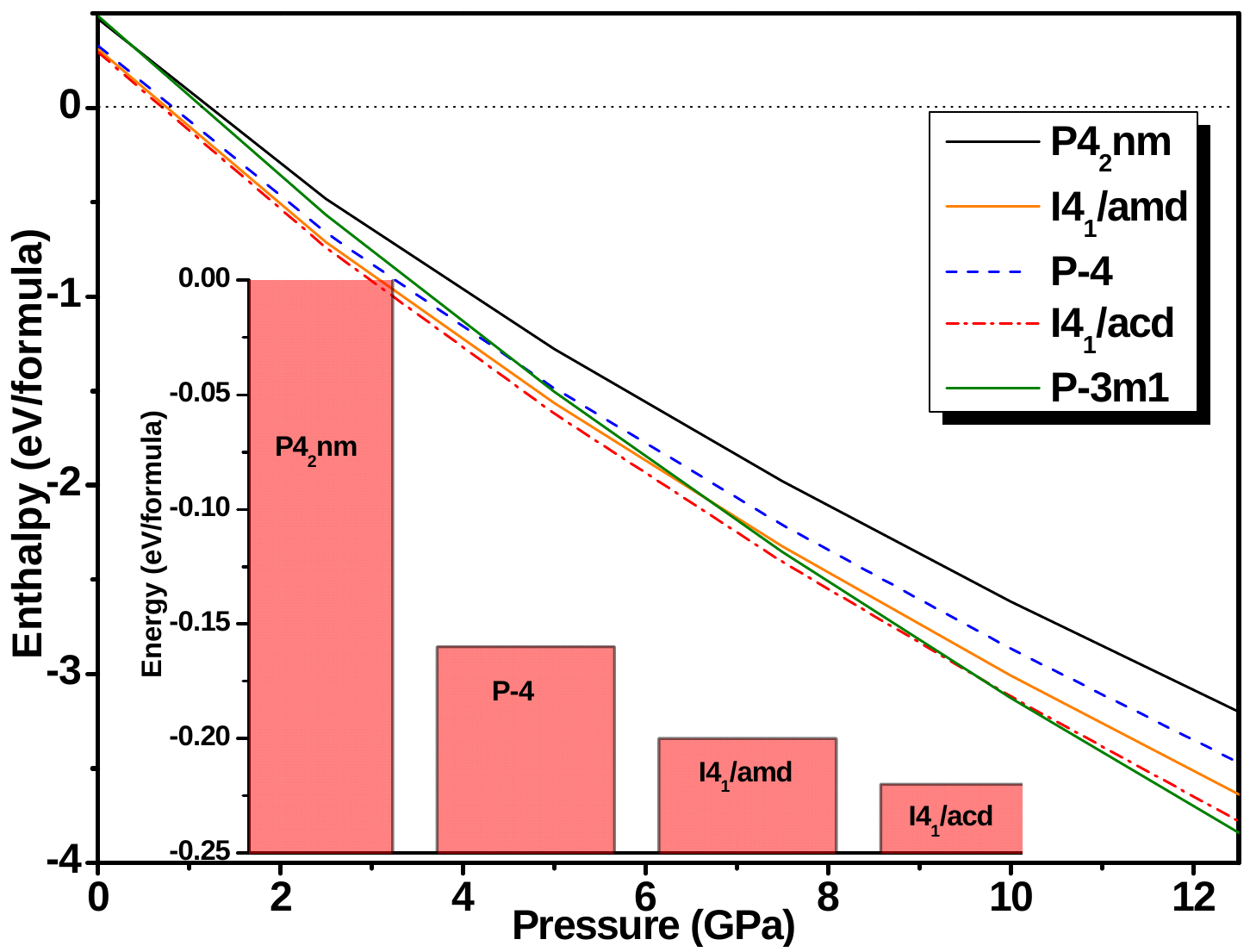}
\caption{%
(Color online) Enthalpy curves (relative to the $\gamma$ phase) of various structures of Mg(BH$_4$)$_2$ as a function of pressure. Enthalpies are given per formula unit. The inset shows the energy per formula unit of $P\bar{4}$, $I4_1/amd$, and $I4_1/acd$ structures including van der Waals interactions at zero pressure relative to the $P4_2nm$ structure.}
\end{center}
\end{figure}

$I4_1/acd$ Mg(BH$_4$)$_2$ becomes more stable than the $\gamma$ phase at pressures above 0.7 GPa (Fig. 2). In the room-temperature experiment, a pressure-induced structural transformation is observed for the porous $\gamma$ phase, and occurs in two steps: The $\gamma$ phase turns into a diffraction-amorphous phase at 0.4-0.9 GPa, and then at approximately 2.1 GPa into the $\delta$ phase \cite{R14}. The calculated phase transition pressure from the $\gamma$ phase to the proposed $\delta$ phase with $P4_2nm$ symmetry is 1.2 GPa (the corresponding phase transition pressure for $P\bar{4}$ phase is 0.8 GPa), which are in good agreement with the experimental values (0.4-0.9 GPa). We note a tiny enthalpy difference between $I4_1/acd$ and $P\bar{4}$ structures at pressures around 1 GPa. As pressure increases to 9.8 GPa, the $P\bar{3}m1$ structure becomes the most stable one, in agreement with earlier predictions \cite{R06,R10}. Bil \textit{et al}. \cite{R08} indicated that it is important to treat long-range dispersion interactions to get the ground state structures of magnesium borohydrides correctly. We have examined the energetic stability of the considered structures through a semiempirical Grimme correction to DFT energies, stresses and forces \cite{R21} (see the inset of Fig. 2). When this correction is included, the $I4_1/acd$ and $P\bar{4}$ structures once again come out as more stable than the $P4_2nm$ structure, by 21.2 kJ/mol and 15.4 kJ/mol, respectively. Energetic stability seems to correlate with the degree of disparity of bond lengths and atomic Bader charges.  The $P4_2nm$ structure has two inequivalent Mg-H distances, 2.26 and 2.07 {\AA}, compared to 2.11 and 2.07 {\AA} in the $I4_1/acd$ structure, and 2.12 and 2.06 {\AA} in the $P\bar{4}$ structure. As we can see, the more homogeneous bond lengths, the greater stability. Bader charges show the same picture: for H atoms, we find them to be -0.63 and -0.59 $e$ in the $P4_2nm$ structure, -0.63 and 0.62 $e$ in the $P\bar{4}$ structure, and -0.63 and -0.61 $e$ in the $I4_1/acd$ structure \cite{R24}. More homogeneous Bader charges and bond lengths in the $I4_1/acd$ and $P\bar{4}$ structures correlate with their greater thermodynamic stability at ambient pressure, in agreement with proposed correlations between local bonding configurations and energetic stability \cite{R13}.

\begin{figure}[h]
\begin{center}
\includegraphics[width=8cm]{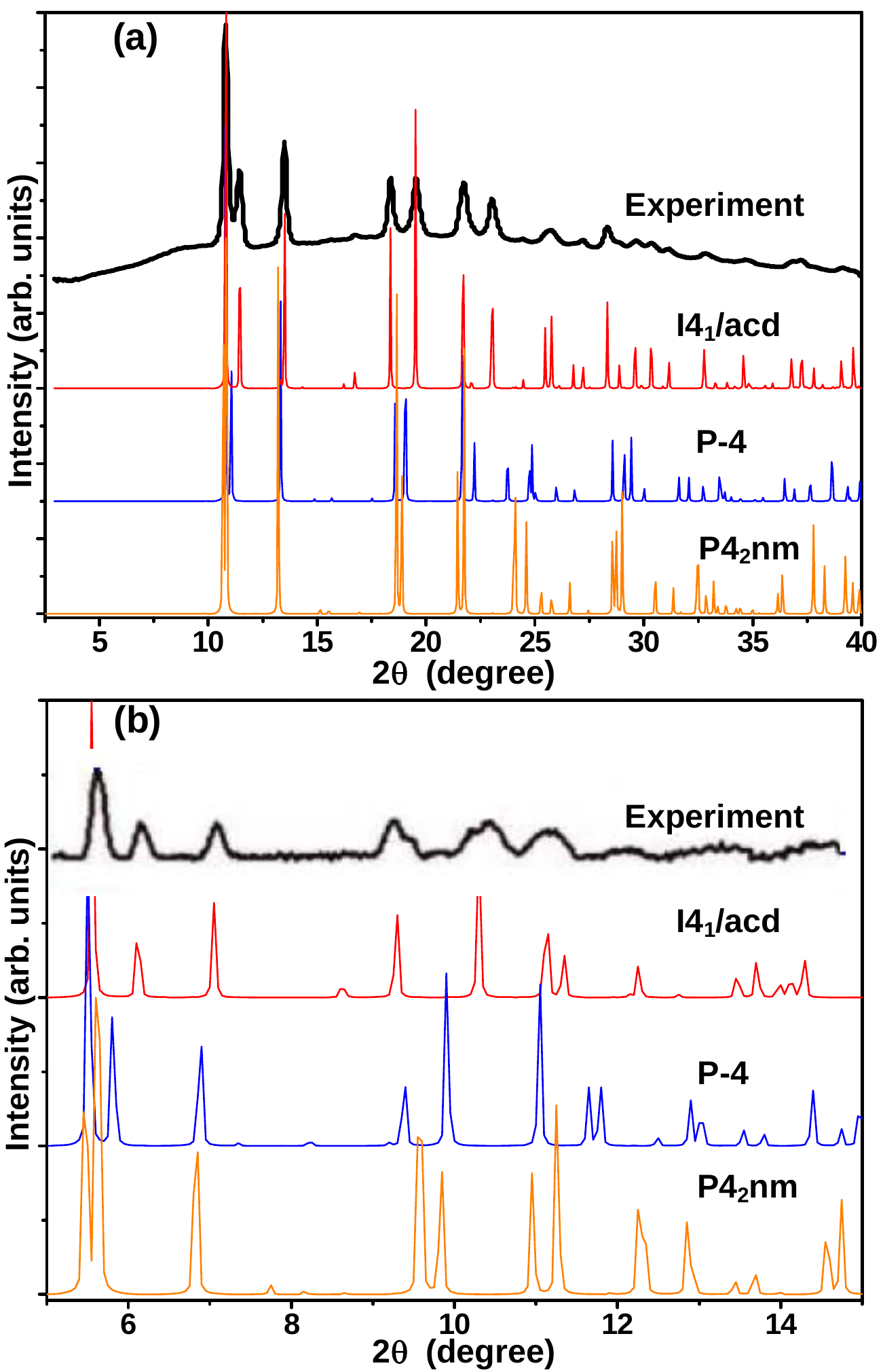}
\caption{%
(Color online) Simulated XRD patterns of the $I4_1/acd$, $P\bar{4}$, and $P4_2nm$ structures of Mg(BH$_4$)$_2$  with the X-ray wavelength of 0.770518 {\AA} at ambient pressure (a) and 0.36814 {\AA} at 10 GPa (b) in comparison with the corresponding experimental results (Ref. 14 and Ref. 6).}
\end{center}
\end{figure}

Our calculations suggest that the $P4_2nm$ structure, proposed by experiment for the $\delta$ phase, is unstable. This implies that either density functional theory calculations are inaccurate for this system, or experimental structure determination was incorrect. To assess these possibilities, we simulated the XRD patterns of the $I4_1/acd$ and $P\bar{4}$ structures, and compared them with the experimental XRD pattern of the $\delta$ phase at ambient pressure (see Fig. 3a). One observes excellent agreement, both for the positions and the intensities of the peaks (including both strong and weak peaks), of the $I4_1/acd$ structure with experiment \cite{R14}. The situation is very peculiar: two structures, $I4_1/acd$ and $P4_2nm$, have nearly identical XRD patterns, both compatible with the experiment -- but one, $I4_1/acd$, is the true thermodynamic ground state (global minimum of the enthalpy), whereas the other, $P4_2nm$, is not even a local minimum of the enthalpy (dynamically unstable structure, incapable of sustaining its own phonons). In this situation, the true structure is clearly $I4_1/acd$. This example gives a clear real-life example of the fact that very different structures can have very similar powder XRD patterns, making structure determination from powder data dangerous, and in such cases input from theory is invaluable. The $P\bar{4}$ structure also has a rather similar XRD pattern, but the peak positions are slightly shifted. Comparison with an independent experimental XRD pattern collected at 10 GPa (Fig. 3b) shows that the peak positions and intensities of the $I4_1/acd$ structure are once again in excellent agreement with the experimental data \cite{R06}, while the strong peaks of the $P\bar{4}$ structure at $9.9^\circ$, $11.6^\circ$, and $11.8^\circ$ obviously deviate from the observed ones. This reinforces our conclusion that the $I4_1/acd$ structure is the best candidate for the high pressure $\delta$ phase. At pressures below 10 GPa a mixture of $I4_1/acd$ and $P\bar{4}$ phases is possible, as the XRD peaks of these two structures are quite similar. We remind that in the experiment, the $\delta$ and $\delta'$ phases are nearly indistinguishable \cite{R14}.

\begin{figure}[h]
\begin{center}
\includegraphics[width=8cm]{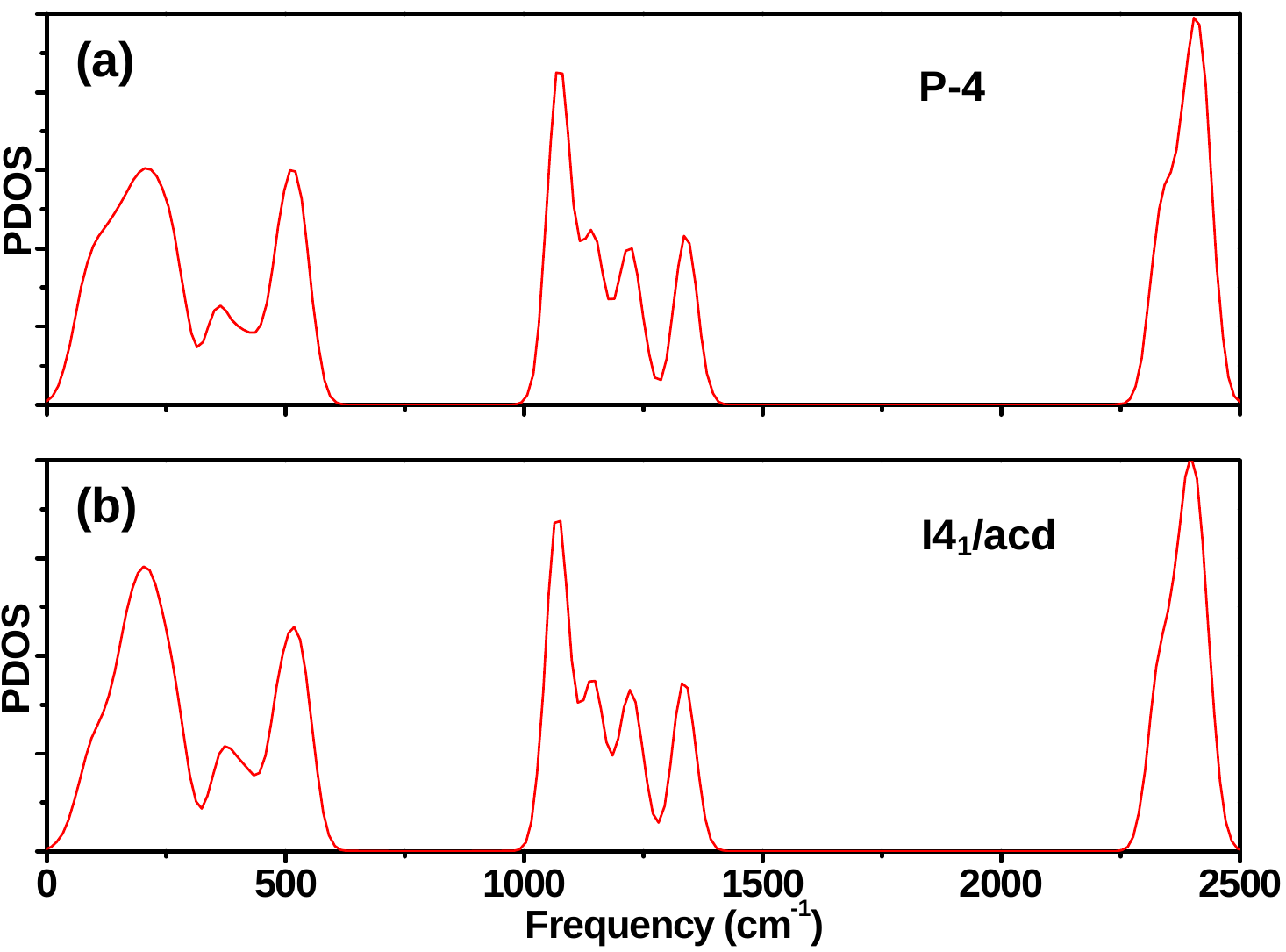}
\caption{%
(Color online) Phonon density of states of (a) the $P\bar{4}$ phase and (b) the $I4_1/acd$ phase at ambient pressure.}
\end{center}
\end{figure}

Filinchuk \textit{et al}. demonstrated the bulk modulus of the $\delta$ phase (28.5 GPa) is almost three times higher than that (10.2 GPa) reported by George \textit{et al}. by fitting the Murnaghan equation of state \cite{R06,R14}. Our third-order Birch-Murnaghan \cite{R25} fits of the equation of state yielded bulk moduli of the $I4_1/acd$ and $P\bar{4}$ structures equal to 24 GPa and 31.7 GPa, respectively, consistent with the measured value (28.5 GPa) \cite{R14}. The observed large density difference with respect to the $\gamma$ phase at ambient conditions (44\%) is equally consistent with 45\% (43\%) for $I4_1/acd$ ($P\bar{4}$) structures \cite{R14}. Therefore, it is difficult to discriminate between the $I4_1/acd$ and $P\bar{4}$ structures by their compression behavior, density or bulk modulus. Our calculations show that the $I4_1/acd$ structure does not only match all experimental observations for the $\delta$ phase and has the lowest enthalpy among all sampled structures at the relevant pressure range, but is also dynamically stable -- phonons were computed at 0, 5 and 10 GPa. The phonon densities of states (PDOS) of $P\bar{4}$ and $I4_1/acd$ phases at ambient pressure are shown in Fig. 4, and once again we see a great degree of similarity. The similarity of all characteristics of these two phases parallels the observed similarity of characteristics of the $\delta$ and $\delta'$ phases and invites one to propose that while the I41/acd structure corresponds to the $\delta$ phase, the $\delta'$ phase may have the $P\bar{4}$ structure.

In conclusion, we performed a systematic structure search for the high pressure phases of magnesium borohydrides and identified two tetragonal structures with space groups $I4_1/acd$ and $P\bar{4}$ as candidates for the unresolved $\delta$ and $\delta'$ phases, respectively, which can be kinetically stable upon the decompression to the ambient conditions. In particular, the $I4_1/acd$ structure is much lower in enthalpy than the earlier reported structures, and the calculated density, bulk modulus, and the simulated XRD patterns are in excellent agreement with the experimental results \cite{R14}. This example highlights the importance of theoretical simulations in establishing crystal structures, when only powder XRD data are available: purely experimental solution may be dangerous even for simple structures, such as the highly symmetric structure of the $\delta$ phase containing only 6 non-hydrogen atoms in the unit cell.

X. F. Zhou thanks Y. Filinchuk for valuable discussions. This work was supported by the National Science Foundation of China (Grant No. 11174152), National 973 Program of China (Grant No. 2012CB921900), the Program for New Century Excellent Talents in University, the Fundamental Research Funds for the Central Universities (Grant No. 65121009), the Postdoctoral Fund of China (Grant No. 201104302), and the open project program of State Key Laboratory of Metastable Materials Science and Technology. A.R.O thanks Intel Corporation, Research Foundation of Stony Brook University, Rosnauka (Russia, contract 02.740.11.5102), DARPA (No. W31P4Q1210008), and National Science Foundation (EAR-1114313) for funding. This work partly used the Extreme Science and Engineering Discovery Environment (XSEDE), which is supported by National Science Foundation grant number OCI-1053575 (Project allocation TG-DMR110058).



\end{document}